\documentclass[aip,jcp,reprint,nofootinbib,citeautoscript,showkeys]{revtex4-1}

\pdfoutput=1
\usepackage{bm}
\usepackage{soul}
\usepackage{braket}
\usepackage{booktabs}
\usepackage{amsmath}
\usepackage{amssymb}
\usepackage{graphicx}
\usepackage{multirow}
\usepackage{tabularx}
\usepackage{hyperref}
\usepackage{xfrac}
\usepackage[usenames,dvipsnames]{xcolor}

\DeclareMathOperator{\Tr}{Tr}
\newcommand{\ci}{\mathrm{i}}
\newcommand{\rmd}{\mathrm{d}}
\DeclareMathOperator{\rmw}{\mathrm{w}}
\DeclareMathOperator{\rmseo}{\mathrm{SEO}}

\DeclareMathOperator{\sigz}{\sigma_{\mathnormal{z}}}
\DeclareMathOperator{\sigxop}{\hat{\sigma}_{\mathnormal{x}}}
\DeclareMathOperator{\sigyop}{\hat{\sigma}_{\mathnormal{y}}}
\DeclareMathOperator{\sigzop}{\hat{\sigma}_{\mathnormal{z}}}

\DeclareMathOperator{\sigzw}{\sigma_{\mathnormal{z}}^{\rmw}}

\newcommand{\mat}[1]{\ensuremath\mathsf{#1}}
\newcommand{\eu}[1]{\ensuremath\mathrm{e}^{#1}}

\newcommand{\Ketbra}[2]{\ensuremath\Ket{#1}\!\Bra{#2}}

\hypersetup{colorlinks=true,linkcolor=blue,citecolor=blue,urlcolor=blue}

\begin{document}

\title{On the identity of the identity operator in nonadiabatic linearized semiclassical dynamics}
\author{Maximilian A.~C.~Saller}
\affiliation{Laboratory of Physical Chemistry, ETH Zurich,
				8093 Zurich, Switzerland}
\author{Aaron Kelly}
\affiliation{Department of Chemistry, Dalhousie University,
				B3H 4R2 Halifax, Canada}
\author{Jeremy O.~Richardson}
\email{jeremy.richardson@phys.chem.ethz.ch}
\affiliation{Laboratory of Physical Chemistry, ETH Zurich,
				8093 Zurich, Switzerland}
\date{\today}

\begin{abstract}
	\noindent
	Simulating the nonadiabatic dynamics of condensed-phase systems
	continues to pose a significant challenge for quantum dynamics methods.
	Approaches based on sampling classical trajectories within the mapping
	formalism, such as the linearized semiclassical	initial value representation
	(LSC-IVR), can be used to approximate quantum correlation functions in
	dissipative environments. Such semiclassical methods however commonly
	fail in quantitatively predicting the electronic-state populations in
	the long-time limit. Here we present a suggestion to minimize this difficulty
	by splitting the problem into two parts, one of which involves the identity,
	and treating this operator by quantum-mechanical principles
	rather than with classical approximations.
	This strategy is applied to numerical
	simulations of spin-boson model systems, showing its potential to drastically
	improve the performance of LSC-IVR and related methods with no change to the
	equations of motion or the algorithm in general, but rather by simply
	using different functional forms of the observables.
\end{abstract}

\maketitle

\section{Introduction}
\label{sec:intro}
	\noindent
	The study of nonadiabatic processes in condensed-phase systems
	remains a challenge for computer simulation.
	These phenomena, occurring when two or more states approach each other
	in energy, influence a wide variety of systems in physics, chemistry and
	biology
	\cite{Marcus1993review,ChandlerET,ConicalIntersections1,Tully2012perspective}.

	Significant efforts continue to be devoted to the development of quantum
	dynamics methods which can accurately capture nonadiabatic effects. While
	grid-based wavefunction propagation shows considerable promise for small or model
	systems \cite{Meyer1990mctdh,Wang2003MLMCTDH,Batista2006mpsoft,
	Richings2015vmcg},
	many realistic problems are simply too large and complex to treat with such
	approaches, given their unfavourable exponential scaling with system size.
	Despite progress in methodology relying, in part, on classical trajectories
	\cite{Martinez1998aims,Saller2015tsa,*Saller2017atsa},
	more approximate semiclassical techniques are
	generally required in the condensed phase
	\cite{Tully1990hopping,Stock1995meanfieldI,Stock2005nonadiabatic,
	Agostini2014classical,Makri2015QCPI,Kapral2015QCL,Tao2018isomorphic}.
	Their favourable, linear scaling with system size allows insight into
	realistic processes to be gained without incurring extreme computational
	costs.

	A typical nonadiabatic process may be described by a continuous nuclear phase
	space and a discrete electronic state space. The total Hamiltonian is given
	by
	\begin{equation}\label{eq:Hgen}
		\hat{H} = \sum_{j=1}^F \frac{\hat{p}_j^{2}}{2m_j} + U(\hat{x}) +
		\hat{V}(\hat{x})\,,
	\end{equation}
	where $x$ and $p$ are $F$-dimensional vectors containing the
	position and momenta for each nuclear degree of freedom with masses $m_j$.
	$U(x)$ is the state-independent potential
	and, for the case of two electronic states,
	the state-dependent potential is
	\begin{equation}
		\hat{V}(x)= V_{1}(x)\Ketbra{1}{1}+V_{2}(x)\Ketbra{2}{2}
		+\Delta(x)\left(\Ketbra{1}{2}\!+\!\Ketbra{2}{1}\right),
	\end{equation}
	where $V_{1}(x)$ and $V_{2}(x)$ are the
	potential-energy surfaces of diabatic states
	$\ket{1}$ and $\ket{2}$
	and $\Delta(x)$ is the coupling between
	the states. While written and presented in the context of two
	electronic states here, all aspects of this work can be generalized to any
	number of electronic states. Reduced units will be employed
	throughout, such that $\hbar=1$.

	The aforementioned issue of mismatched discrete and continuous spaces clearly
	arises from the final term in Eq.~\ref{eq:Hgen}\@. In order to express
	the problem in terms of continuous variables, the electronic states can
	be mapped onto a space of singly-excited harmonic oscillators (SEOs)
	\cite{Stock1997mapping}.
	Projecting into the space of the SEOs yields the
	wavefunction
	\begin{equation}\label{eq:map}
		\braket{\mat{X}|n} = \sqrt\frac{2}{\pi} X_n \, \eu{-(X_1^2+X_2^2)/2}
	\end{equation}
	in terms of positions, $\mat{X}$, or equivalently momenta, $\mat{P}$,
	which are 2-dimensional vectors known as mapping variables.
	This allows the entire system to be described \textit{via} an	extended phase
	space, $\{x,p,\mat{X},\mat{P}\}$, which can be sampled and explored using
	classical trajectories. \cite{Meyer1979nonadiabatic,Kelly2012mapping}
	The classical Hamiltonian underlying the mapping approach is given by
	\begin{equation}\label{eq:mapHam}
		\mathcal{H} = \sum_{j=1}^F \frac{p_j^{2}}{2m_j} + U(x) +
		\frac{1}{2}\left[\mat{X}^{\mathrm{T}}\mat{V}\mat{X}+
		\mat{P}^{\mathrm{T}}\mat{V}\mat{P}-
		\mathrm{tr}\,\mat{V}\right],
	\end{equation}
	where $\mat{V}$ is the matrix representation of	$\hat{V}(x)$.

	Owing to the relative simplicity of the mapping approach and the convenience
	of being able to describe the total system in terms of an extended phase space
	that grows only linearly with the number of nuclear degrees of freedom and
	electronic states,
	a considerable number
	of approaches have been developed with it at their core
	\cite{Meyer1979nonadiabatic,Sun1997mapping,Sun1998mapping,Wang1999mapping,
	Kim2008Liouville,Hsieh2012FBTS,*Hsieh2013FBTS,
	Kelly2012mapping,Mueller1998mapping,
	Bonella2005LANDmap,Huo2011densitymatrix,Ananth2010mapping,mapping,vibronic,
	Ananth2013MVRPMD,Duke2015MVRPMD,Hele2016Faraday,Chowdhury2017CSRPMD,
	Bonella2003mapping,Cotton2013mapping,*Miller2016Faraday,Stock1997mapping,
	Thoss1999mapping,*Thoss2000pyrazine,Church2018MQCIVR}.

	Many semiclassical nonadiabatic simulation methods
	struggle in predicting the correct long-time limit of observables for
	asymmetric systems.
	One suggestion for improving the accuracy has been to
	utilize the formalism of master equations \cite{Shi2003master}.
	The benefits of this include a noticeable gain in accuracy as
	well as the ability to obtain long time dynamics using only short time
	trajectories
	\cite{Kelly2015nonadiabatic,Kelly2016master,Reichman2016spinboson,
	Reichman2017spinboson}.
	A similar motivation underlies the approach presented here, albeit using a
	simpler strategy.
	First we recast the observable of interest in terms of a correlation function
	involving the identity operator.
	In quantum mechanics, the effect of the identity operator
	is known to be exactly unity. The classical equivalent of
	this operation is clearly to multiply by the number one.
	We therefore calculate the correlation function
	with this definition of the identity, instead of using the
	approximation provided by the semiclassical trajectories.
	The result is a marked improvement in accuracy.

\section{Theory}
\label{sec:theo}
	\noindent
	In this section we present the formalism for evaluating correlation functions
	using semiclassical methods and consider the various options for classical
	representations of the quantum operators.

\subsection{Linearized semiclassical initial value representation}
\label{sec:lscivr}
	\noindent
	In the study of condensed-phase systems, the quantity of interest can
	typically be expressed as a time correlation function between two quantum operators,
	$\hat{A}$ and $\hat{B}$, given by
	\begin{equation}\label{eq:Cexact}
		C_{AB}(t) = \Tr\left[ \hat{\rho} \hat{A} \, \eu{\ci\hat{H}t}
					\hat{B} \, \eu{-\ci\hat{H}t} \right],
	\end{equation}
	where $\hat{\rho}$ is the initial density matrix, normalized such that
	$\Tr[\hat{\rho}]=1$.

	In order to take the semiclassical approximation, it is necessary to rewrite
	the correlation function in terms of the mapping variables. This is
	readily achieved using the Wigner transform, given, for any general quantum
	operator, $\mathcal{\hat{O}}$, by
	\begin{eqnarray}\label{eq:Wtransf}
		\mathcal{O}^{\rmw} (x,p,\mat{X},\mat{P}) =
		\qquad\qquad\qquad\qquad\nonumber\\
		\!\!\!\iint\!\! \eu{\ci p\cdot y + \ci \mat{P}\cdot \mat{Y}}
		\!\!\Braket{\! x-\frac{y}{2}, \mat{X}-\vspace{1em}\frac{\mat{Y}}{2}\! |\!
		\hat{\mathcal{O}}\! |\! x+\frac{y}{2}, \mat{X}+\frac{\mat{Y}}{2} }
		\rmd y \, \rmd \mat{Y}\!,
	\end{eqnarray}
	where $y$ and $\mathsf{Y}$ are dummy variables of the same shape as $x$
	and $\mathsf{X}$. Using this phase-space representation,
	the exact quantum correlation function can be written as
	\begin{eqnarray}\label{eq:Clscivr_ex}
		C_{AB}(t) = \frac{1}{(2\pi)^{F+L}} \iiiint
		\left[\hat{\rho}
		\hat{A}\right]^{\rmw}(x,p,\mat{X},\mat{P})\qquad\nonumber\\
		\left[\hat{B}(t)\right]^{\rmw}(x,p,\mat{X},\mat{P})
		\, \rmd x \, \rmd p \, \rmd\mat{X} \, \rmd\mat{P}\,.
	\end{eqnarray}
	No approximation has been made, and in this expression the time-dependent operator
	$\hat{B}(t)\equiv\eu{\ci\hat{H}t}\hat{B}\eu{-\ci\hat{H}t}$ has to be
	computed by quantum-mechanical propagation. In cases, such as the ones
	studied here, where the density matrix, $\hat{\rho}$, and the operator
	$\hat{A}$ commute, the Wigner transform of the product,
	$[\hat{\rho}\hat{A}]^{\rmw}(x,p,\mat{X},\mat{P})\equiv\rho^{\rmw}
	(x,p,\mat{X},\mat{P})A^{\rmw} (x,p,\mat{X},\mat{P})$, may be used instead.

	An approximation to the correlation function is obtained if classical
	trajectories defined by the mapping Hamiltonian of Eq.~\ref{eq:mapHam}, are
	employed to determine the value of the Wigner-transformed operator at time
	$t$ \cite{Miller2001SCIVR}.
	This gives
	\begin{eqnarray}\label{eq:Clscivr}
		C_{AB}(t) \approx \frac{1}{(2\pi)^{F+L}} \iiiint
		\rho^{\rmw} (x,p,\mat{X},\mat{P})
		A^{\rmw} (x,p,\mat{X},\mat{P})\nonumber\\
		B^{\rmw} (x(t),p(t),\mat{X}(t),\mat{P}(t))
		\,\rmd x \, \rmd p \, \rmd\mat{X} \, \rmd\mat{P}\,.\qquad
	\end{eqnarray}
	This expression is known in the literature as the linearized semiclassical
	initial value representation (LSC-IVR) \cite{Sun1998mapping}. Notably,
	the approximation in Eq.~\ref{eq:Clscivr} is exact at $t=0$ but depending on
	the problem may give poor results in the long-time limit.

\subsection{Pauli spin operators}
\label{sec:Pauli}
	\noindent
	Many nonadiabatic processes of interest occur between two electronic states.
	This includes for example electron-transfer reactions
	\cite{Marcus1993review,ChandlerET}
	and dynamics near conical intersections \cite{ConicalIntersections1}.
	In general, all operators relevant to a two-state quantum system can be
	written in terms of the traceless Pauli
	spin matrices, $\sigxop$, $\sigyop$ and $\sigzop$, and the identity,
	$\hat{\mathcal{I}}$. Employing the mapping	approach allows these operators
	to be expressed as \cite{Stock1997mapping}
	\begin{subequations}\label{eq:sigmas}
		\begin{align}
			\sigxop &= \Ketbra{2}{1}+\Ketbra{1}{2}
			\mapsto \hat{X}_{1}\hat{X}_{2}+\hat{P}_{1}\hat{P}_2\\
			\sigyop &= \ci\left(\Ketbra{2}{1}-\Ketbra{1}{2}\right)
			\mapsto\hat{X}_{1}\hat{P}_{2}-\hat{P}_{1}\hat{X}_{2}\\
			\sigzop &= \Ketbra{1}{1}-\Ketbra{2}{2}
			\!\mapsto\!\tfrac{1}{2}\!\!\left(\hat{X}_{1}^{2}+\hat{P}_{1}^{2}-
			\hat{X}_{2}^{2}-\hat{P}_{2}^{2}\right)\\
			\hat{\mathcal{I}} &= \Ketbra{1}{1}+\Ketbra{2}{2}
			\!\mapsto\!\tfrac{1}{2}\!\!\left(\hat{X}_{1}^{2}\!+\!\hat{P}_{1}^{2}+
			\hat{X}_{2}^{2}+\hat{P}_{2}^{2}-2\right)\label{eq:sig0}\!.
		\end{align}
	\end{subequations}
	In order to calculate LSC-IVR correlation functions \textit{via}
	Eq.~\ref{eq:Clscivr}, these operators must be Wigner transformed to yield
	phase-space representations in terms of the mapping variables, $\mat{X}$ and
	$\mat{P}$, which can be evolved with classical trajectories.

	For each of these operators, there are two possible forms of the Wigner
	transform, one which is projected onto the SEO subspace, and one which is not
	\cite{Kelly2012mapping}. Wigner transforming the operators in the form given
	on the right-hand sides of Eqs.~\ref{eq:sigmas}, \textit{i.e.}~without
	projecting into the SEO subspace, yields phase-space representations of
	the same functional form in terms of the mapping variables,
	\textit{e.g.}
	\begin{subequations}\label{eq:opclass}
	\begin{align}
		\sigzw(\mat{X},\mat{P}) &= \tfrac{1}{2}\left(X_1^2 + P_1^2 -
		X_2^2 - P_2^2\right)\\
		\mathcal{I}^{\rmw}(\mat{X},\mat{P}) &=\tfrac{1}{2}\left(X_1^2 + P_1^2
		+X_2^2 + P_2^2 - 2\right).
	\end{align}
	\end{subequations}
	On the other hand, taking the Wigner transform of the state-state
	representation and using the SEO projection, Eq.~\ref{eq:map}, yields
	\begin{subequations}\label{eq:opwig}
	\begin{align}
		\sigma_{\alpha}^{\rmseo}\left(\mat{X},\mat{P}\right) &=
		\sigma_{\alpha}^{\rmw} \phi
		\qquad \alpha \in \left\{x,y,z\right\}\\
		\mathcal{I}^{\rmseo}(\mat{X},\mat{P}) &=\tilde{\mathcal{I}}\phi\,,
	\end{align}
	\end{subequations}
	where
	\begin{equation}
		\tilde{\mathcal{I}}=\tfrac{1}{2} \left(
		X_{1}^{2}+P_{1}^{2}+X_{2}^{2}+P_{2}^{2} - 1\right),
	\end{equation}
	and
	\begin{align}
		\phi &=16 \exp\left(-X_{1}^{2}-X_{2}^{2}-P_{1}^{2}-P_{2}^{2}\right).
	\end{align}
	Note that unlike for the Pauli spin operators, the projected Wigner function
	$\mathcal{I}^{\rmseo}$ is not simply equal to its unprojected form,
	$\mathcal{I}^{\rmw}$, multiplied by $\phi$.

\subsection{Observables and Correlation functions}
\label{sec:Corrs}
	\noindent
 	In two-state quantum systems, a quantity of particular interest is the
	population difference between the	two	states, $P(t)$, measured
	by the operator $\sigzop$. A commonly studied
	\cite{Makri1992QUAPI,Kelly2015nonadiabatic,Kelly2016master,Reichman2016spinboson},
	non-equilibrium initial condition is a separable bath thermal density with
	the system purely in one electronic state,
	\textit{i.e.}~$\hat{\rho}_\text{b}\Ketbra{1}{1}$. This is normalized such that
	its trace is unity.

	In this case, the population difference is given by
	\begin{subequations}
	\begin{align}
		P(t) &= \mathrm{Tr}[\hat{\rho}_\mathrm{b}
		\Ketbra{1}{1}\sigzop(t)]\label{eq:Pstate}\\
		&= C_{\mathcal{I}\sigz}(t) + C_{\sigz\!\sigz}(t)\,,
		\label{eq:Pop}
	\end{align}
	\end{subequations}
	where the normalized density matrix appearing in the correlation functions is
	$\hat{\rho}=\hat{\rho}_{\mathrm{b}}/2$.
	These two expressions are identical because
	$(\hat{\mathcal{I}}+\sigzop)/2=\Ketbra{1}{1}$,
	and we have simply separated the operator into trace-containing and traceless parts
	for clarity in what follows.

	By considering known relations of the Pauli matrices, one can show that
	$C_{\sigz\!\sigz}(t)$ has an initial value of one and decays
	to zero in the long-time limit. $C_{\mathcal{I}\sigz}(t)$ on the other hand,
	starts at zero and plateaus at long times to a system-dependent finite value.
	Notably most semiclassical approximations, including
	LSC-IVR, capture the short time behaviour of these two correlation functions
	and for symmetry reasons the long-time decay of $C_{\sigz\!\sigz}(t)$ to zero
	as well. It is however much more difficult to obtain an accurate quantitative
	prediction of the long-time limit of $C_{\mathcal{I}\sigz}(t)$.

	As a result of the different forms of the Wigner functions, there are number of
	ways in which the LSC-IVR correlation function could be evaluated,
	and in general they will give different approximations.
	Note that in order to ensure that the mapping variables describe the physical
	system, it is necessary for at least one of the operators to be projected
	onto the SEO subspace.

	As shown in Eq.~\ref{eq:sig0} there exists an expression for the identity
	operator in terms of the mapping variables. According to quantum mechanics, this
	operator measures the total population of the system and should always return
	exactly unity. However, it is known that the classical trajectories do not obey
	this rule correctly, and this is one of the main sources of error in the
	approximation \cite{Mueller1998mapping,Kelly2012mapping,Bonella2003mapping}.
	We propose that some of the errors associated with this can be
	eliminated by invoking the exact nature of the identity operator. In practice
	this is achieved by simply replacing the mapping representation of the
	identity with the number 1 when calculating correlation functions.

	\begin{table}
		\caption{Different formulations of the LSC-IVR
		correlation function, as shown in Eq.~\ref{eq:Clscivr}.
		We define $\rho^{\rmw}=\rho_{\mathrm{b}}^{\rmw}\rho_{\mathrm{el}}^{\rmw}/2$.
		\label{tab:Clscivr}}
		\setlength{\tabcolsep}{8pt}
		\begin{tabular}{c|c|cc|cc}
			\hline
			\hline
			\multirow{2}{*}{method} & \multirow{2}{*}{$\rho^{\rmw}_{\mathrm{el}}(\mat{X},\mat{P})$} &
			\multicolumn{2}{c|}{$C_{\mathcal{I}\sigz}(t)$} &
			\multicolumn{2}{c}{$C_{\sigz\!\sigz}(t)$} \\
			\cline{3-6}
			& & $A^{\rmw}$ & $B^{\rmw}$ & $A^{\rmw}$ & $B^{\rmw}$ \\
			\hline
			``double unity'' & $\phi^{2}$ & 1 & $\sigzw$ & $\sigzw$ & $\sigzw$ \\
			``double SEO'' & $\phi^{2}$ & $\tilde{\mathcal{I}}$ & $\sigzw$ &
			$\sigzw$ & $\sigzw$ \\
			``single unity'' & $\phi$ & $1$ & $\sigzw$ & $\sigzw$ & $\sigzw$ \\
			``single Wigner'' & $\phi$ & $\mathcal{I}^{\rmw}$ & $\sigzw$ & $\sigzw$ & $\sigzw$\\
			``single SEO'' & $\phi$ & $\tilde{\mathcal{I}}$ & $\sigzw$ & $\sigzw$ &	$\sigzw$ \\
			\hline
			\hline
		\end{tabular}
	\end{table}

	Table \ref{tab:Clscivr} shows the different formulations of the LSC-IVR
	correlation functions that are employed here. The terms ``single" and
	``double" refer to whether only one, or both operators are projected into the
	SEO subspace. ``Unity", ``SEO" and
	``Wigner" are used to identify what functional form of the identity
	operator is used. There are thus five distinct LSC-IVR approximations for
	$C_{\mathcal{I}\sigz}(t)$. However, there are only two possibilities for
	$C_{\sigz\!\sigz}(t)$, using either ``single'' or ``double'' projections.
	Note that, as the function $\phi$ is time-independent under dynamics of
	$\mathcal{H}$, we have chosen to include it explicitly in the formulation
	of the density matrix,
	$\rho^{\rmw}=\rho_{\mathrm{b}}^{\rmw}\rho_{\mathrm{el}}^{\rmw}/2$,
	which defines the sampling distribution of initial conditions. \cite{Sun1998mapping}

	Our approach is therefore to compute $P(t)$ via two correlation functions,
	as defined in Eq.~\ref{eq:Pop}. We employ the LSC-IVR approximation,
	Eq.~\ref{eq:Clscivr}, for each correlation function using the functions
	defined in Table~\ref{tab:Clscivr}\@. Both correlation
	functions can be evaluated by a single simulation, and in fact since the
	$B^{\rmw}$ functions are the same in each case, one could equivalently
	add the two $A^{\rmw}$
	functions together to give a single correlation function. The standard
	LSC-IVR approach advocated in Ref.~\onlinecite{Sun1998mapping} is equivalent
	to the ``double SEO'' approach. It uses
	$A^{\rmw}=\tilde{\mathcal{I}}+\sigzw=X_1^2+P_1^2-\tfrac{1}{2}$,
	$B^{\rmw}=\sigzw$ and $\rho^{\rmw}_{\mathrm{el}}=\phi^2$.
	In our tests, this is actually one of the least accurate of the approaches
	defined in the table. The Poisson-bracket mapping equation (PBME)
	approach described in Ref.~\onlinecite{Kim2008Liouville} on the other
	hand corresponds to ``single SEO'', which is the same
	except for the sampling distribution $\rho^{\rmw}_{\mathrm{el}}=\phi$.
	The formulation which we find to be most accurate is ``double unity'',
	which is equivalent to using $A^{\rmw}=1+\sigzw$
	with $\rho^{\rmw}_{\mathrm{el}}=\phi^{2}$.

\section{Results and Discussion}
\label{sec:rnd}
	\noindent
	In order to test the accuracy of the various approaches proposed here,
	we perform numerical simulations on the spin-boson model and compare the
	predicted time-dependent population difference with exact results.

\subsection{The spin-boson Hamiltonian}
\label{sec:SBHam}
	\noindent
	The spin-boson Hamiltonian serves as a simple, yet challenging model which
	incorporates the key aspects of dissipative quantum systems
	\cite{Garg1985spinboson,Leggett1987spinboson,Weiss}. Owing to the wide variety of dynamical
	regimes accessible \textit{via} its relatively few parameters, it has been
	thoroughly studied, and exact results are available
	\cite{Mak1991spinboson,Makarov1993nonadiabatic,Topaler1996nonadiabatic,
	Thoss2001hybrid} to serve as a benchmark for new
	quantum dynamics methods
	\cite{Wang1999mapping,Kim2008Liouville,Hsieh2012FBTS,*Hsieh2013FBTS,
	Bonella2005LANDmap,Kelly2015nonadiabatic,Kelly2016master,Cotton2013mapping,*Miller2016Faraday,
	Shi2003master,Reichman2016spinboson,Reichman2017spinboson,Miller2001SCIVR}.

	The spin-boson model is a two-state quantum system, coupled to a bath of
	harmonic oscillators. In the context of the general
	Hamiltonian given in Eq.~\ref{eq:Hgen}, it is defined by
	\begin{subequations}\label{eq:SBham}
	\begin{align}
		\hat{V}(x)&=\varepsilon\sigzop+\Delta\sigxop+
		\sum\limits_{j=1}^{F} c_{j}x_{j}\sigzop\\
		U(x)&=\frac{1}{2}\sum\limits_{j=1}^{F} m_j \omega_{j}^{2}x_{j}^2\,,
	\end{align}
	\end{subequations}
	where $\varepsilon$ and $\Delta$ are the energy bias and constant diabatic coupling
	between the two electronic states forming the quantum system.
	The $j$th nuclear degree of freedom has frequency $\omega_{j}$ and vibronic coupling coefficient $c_j$.
	Note that the classical equations of motion are defined
	by the Hamiltonian, Eq.~\ref{eq:mapHam},
	using the traceless
	form of the $\hat{V}(x)$ operator, as shown above.

	All the properties of the bath are described entirely by the spectral density.
	The two most commonly studied spectral densities are the Ohmic and Debye
	forms, $J_{\mathrm{Oh}}(\omega)$ and $J_{\mathrm{De}}(\omega)$, given by
	\begin{subequations}
		\begin{align}
			J_{\mathrm{Oh}}(\omega) &= \eta \omega \exp\left(-\frac{\omega}
			{\omega_{c}}\right)\label{eq:JOhmic}\\
			J_{\mathrm{De}}(\omega) &= \eta\omega\,\frac{\omega_{c}}
			{\omega^{2}+\omega_{c}^{2}}\label{eq:JDebye}\,,
		\end{align}
	\end{subequations}
	where $\omega_{c}$ is the characteristic frequency of the bath and $\eta$
	is the coupling strength. In the Ohmic case, the Kondo parameter,
	$\xi=2\eta/\pi$, is often used to give the strength of the system-bath
	coupling instead.
	It is worth noting that
	the Debye spectral density spans a frequency range much broader than that of
	$J_{\mathrm{Oh}}(\omega)$ and, as a result, constitutes a greater numerical
	challenge \cite{Wang1999mapping}.

	In order to perform numerical simulations it is necessary to discretize the
	bath into the form
	\begin{equation}\label{eq:Jexact}
		J(\omega) = \frac{\pi}{2}\sum\limits_{j=1}^F
		\frac{c_{j}^{2}}{m_j\omega_{j}} \delta(\omega - \omega_{j})\,,
	\end{equation}
	for which a number of such schemes have been proposed
	\cite{Mueller1999pyrazine,Wang2001hybrid,Craig2007condensed,RPMDrate}. In
	this work, a strategy which reproduces exact values for the reorganization
	energy was used \cite{Wang2001hybrid,Craig2007condensed}. The choice of mass, $m_j$,
	has no effect on the results.

	A set of simulations, calculating the population difference between the two
	quantum states of the spin-boson Hamiltonian were carried out, using both
	Ohmic and Debye spectral densities. In each case the
	calculations were converged with respect to the number of nuclear degrees
	of freedom used for the bath discretization.

	The phase-space integrals in Eq.~\ref{eq:Clscivr} were carried out by Monte
	Carlo importance sampling. Initial conditions for the nuclear modes were drawn from
	the thermal Wigner distribution of the uncoupled bath,
	\begin{eqnarray}
		\rho^{\rmw}_{\mathrm{b}}(x,p) = \prod_{j=1}^{F}
		2\tanh\left(\tfrac{1}{2}\beta\omega_{j}\right)
		\qquad\qquad\qquad\qquad\nonumber\\
		\exp\left[
		-\left(\frac{p_{j}^{2}}{m_j\omega_j}+m_{j}\omega_{j}x_{j}^{2}\right)
		\tanh\left(\tfrac{1}{2}\beta\omega_{j}\right)\right],
	\end{eqnarray}
	where $\beta$ is the inverse temperature. $\rho^{\rmw}_{\mathrm{el}}$
	provides a convenient distribution from which to sample the initial conditions for the
	mapping variables \cite{Sun1998mapping}.

	The results of the various definitions of the LSC-IVR correlation functions
	are compared with
	numerically exact quasiadiabatic propagator
	path-integral (QUAPI) calculations \cite{Makri1992QUAPI}.

\subsection{Ohmic bath}

	\begin{figure}[t!]
		\centering
		\includegraphics[width=3.37in]{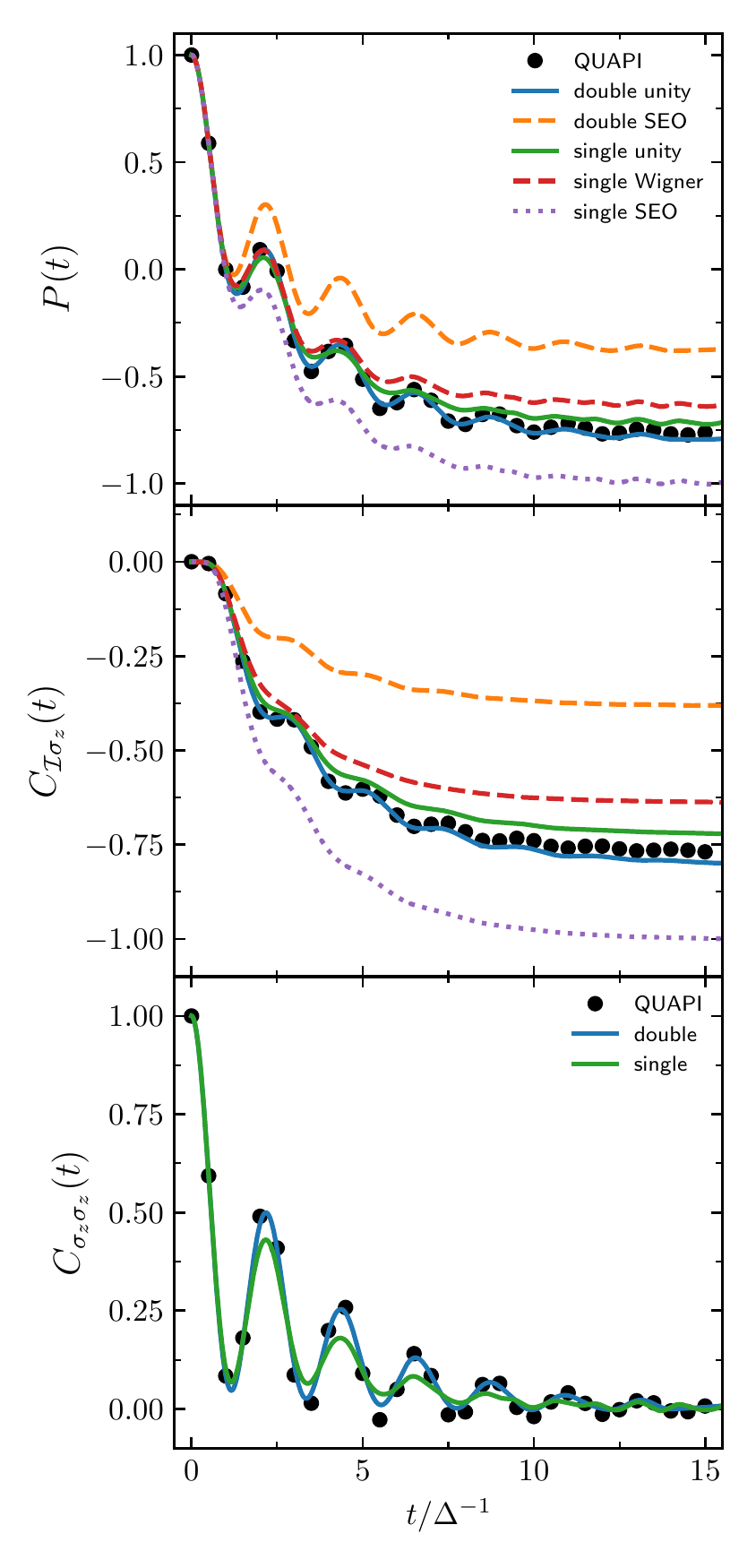}
		\caption{Time-dependent population difference and the constituent
		correlation functions obtained using
		different LSC-IVR definitions for a spin-boson model
		with an Ohmic bath and parameters $\varepsilon=\Delta$,
		$\beta=10\Delta^{-1}$, $\omega_c=2.5\Delta$ and $\xi=0.2$. The various
		approaches are defined in Table~\ref{tab:Clscivr}.
		\label{fig:Ohmic1}}
	\end{figure}

	\noindent
	The first system chosen corresponds to the intermediate regime between
	strongly incoherent decay and coherent oscillations and is characterized
	by moderate system-bath coupling.
	The	particular set of parameters has been extensively
	studied using a number of different semiclassical dynamics methods
	\cite{Kelly2015nonadiabatic,Kelly2016master}. Here, simulations were found to converge
	with a total of $10^{6}$ trajectories,
	using a bath of $F=36$ nuclear degrees of freedom and employing a timestep of
	$\delta t=0.01\Delta^{-1}$.

	Fig.~\ref{fig:Ohmic1} shows the time-dependent population difference, $P(t)$,
	as well as the correlation functions from which it was constructed, in accordance
	with the definitions given in Table~\ref{tab:Clscivr}\@. Notably, both ``single SEO''
	and ``double SEO'' fail to capture the correct asymptotic limit of $P(t)$.
	Closer inspection of the constituent correlation functions, $C_{\sigz\!\sigz}(t)$
	and $C_{\mathcal{I}\sigz}(t)$, reveals that while the ``double'' approach of sampling
	electronic initial conditions from $\phi^{2}$ performs somewhat better than
	single'' for $C_{\sigz\!\sigz}(t)$,
	any significant errors in $P(t)$ arise from approximations to $C_{\mathcal{I}\sigz}(t)$.
	The ``single Wigner'' result is an improvement over using the SEO operator,
	although
	employing our approach of setting the identity
	operator to the number 1, is clearly the most accurate of all.
	Both the ``single unity'' and
	``double unity'' formulation of the $C_{\mathcal{I}\sigz}(t)$ correlation function
	capture the long-time limit of its decay.
	As a result the population differences resulting from these two
	methods yield a drastic improvement over all the other results,
	with ``double unity'' in particular yielding close to quantitative accuracy.

	To ascertain the general validity of our approach, a second parameter
	set was investigated for the Ohmic spectral density. Previously studied
	with a number of methods,\cite{Kelly2015nonadiabatic} this system is characterized
	by stronger system-bath coupling, which in this case results in critical damping.
	The timestep and convergence parameters were identical to those of the first system.

	\begin{figure}[t!]
		\includegraphics[width=3.37in]{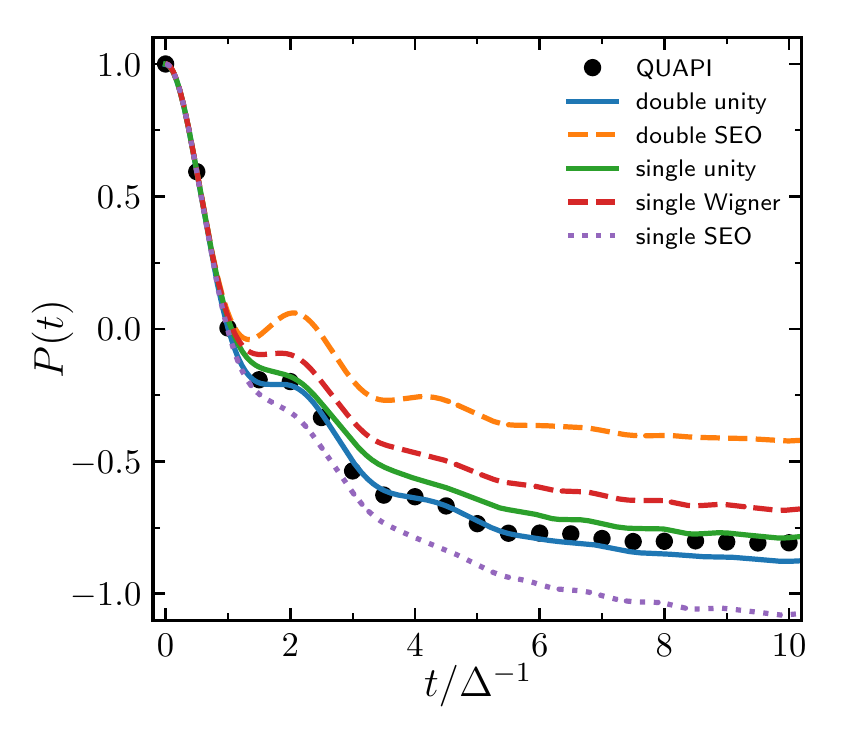}
		\caption{As top panel of Fig.~\ref{fig:Ohmic1} with parameters
		$\varepsilon=\Delta$, $\beta=5\Delta^{-1}$, $\omega_c=2\Delta$ and
		$\xi=0.4$.
		\label{fig:Ohmic2}}
	\end{figure}

	Fig.~\ref{fig:Ohmic2} shows the population differences, again calculated using the
	approaches outlined in Table \ref{tab:Clscivr}, for this second parameter set.
	The accuracy of the different approaches with respect to the exact QUAPI reference result
	is very similar to that shown in Fig.~\ref{fig:Ohmic1}\@. ``Single SEO'' and ``double
	SEO'' fail to reach the correct long-time asymptote
	of the population difference. In addition, with the stronger system-bath coupling present,
	``double SEO'' reports spurious oscillatory structure in the short time limit.
	It would therefore fail to identify this particular parameter set as resulting
	in critical damping.
	Again ``single Wigner'' yields somewhat better results, and
	both ``single unity'' and ``double unity'', corresponding to our new strategy
	of setting the identity equal to 1, considerably outperform all other approaches.

\subsection{Debye bath}

	\begin{figure}[t!]
		\includegraphics[width=3.37in]{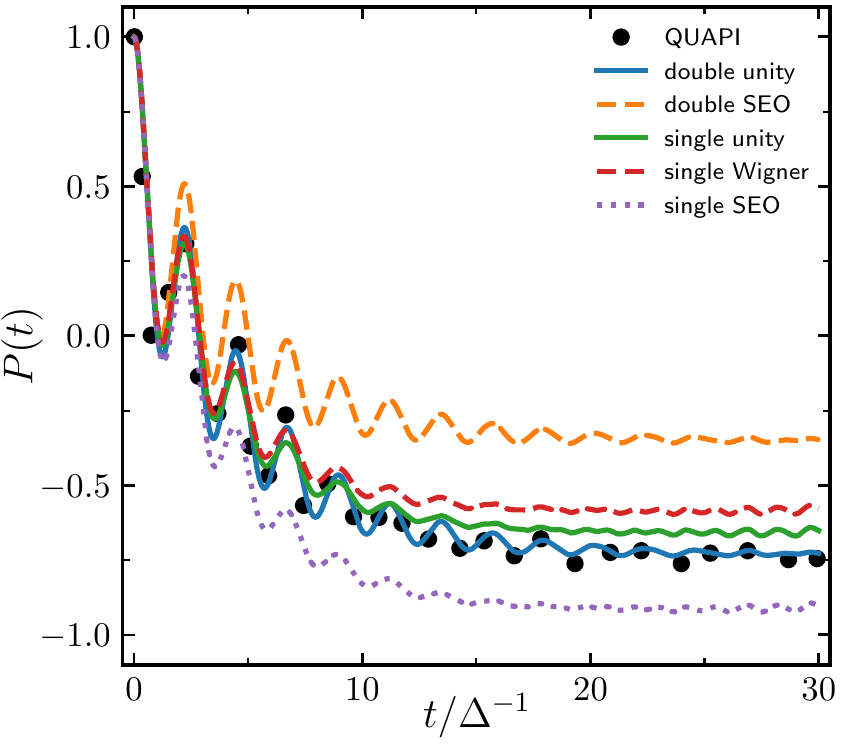}
		\caption{As top panel of Fig.~\ref{fig:Ohmic1} but with a Debye bath and parameters
		$\varepsilon=\Delta$, $\beta=50\Delta^{-1}$, $\omega_c=5\Delta$ and
		$\eta=0.25$. Exact data taken from
		Ref~\onlinecite{Reichman2016spinboson}.
		\label{fig:Debye1}}
	\end{figure}

	\noindent
	As mentioned in Sec.~\ref{sec:SBHam}, the Debye spectral density is
	significantly more challenging on account of spanning a broader range of
	frequencies. Consequently, with the parameters chosen here, results were
	found to converge using a bath of $F=60$ nuclear degrees of freedom,
	averaging over a total of $10^{6}$ semiclassical trajectories with a timestep of
	$\delta t=0.0025\Delta^{-1}$. The shorter timestep was required in order to
	accurately treat the higher frequencies contained in the bath.

	The set of parameters chosen represent the most coherent of the three systems
	reported here, with the weakest system-bath coupling.
	Results for the time-dependent population difference are shown in
	Fig.~\ref{fig:Debye1}.
	Again, similarly to both Ohmic systems discussed above,
	the ``single SEO'' and ``double SEO'' approaches fail to capture the
	correct long-time limit of the population difference. ``Single Wigner''
	somewhat improves on this, but our new approach, ``double unity,'' yields
	the best results by a significant margin,
	approaching quantitative accuracy with respect to the QUAPI benchmark.

\section{Conclusion}
\label{sec:conc}
	\noindent
	Based on our study, we propose a simple modification of the standard
	nonadiabatic LSC-IVR approximation for quantum correlation functions.
	Wherever possible, the correlation function of interest should be rewritten
	in terms of Pauli matrices and the identity,
	and given that the effect of the latter is known to be
	exactly unity, it should not be estimated by the classical mapping variables,
	but replaced by its known value, the number 1.
	Although we have only presented a formulation for
	dynamical simulation of a two-state system, the mapping approach
	can be extended to an arbitrary	number of electronic
	states.\cite{Meyer1979nonadiabatic,Stock1997mapping}
	Also, as it is possible to split any Hermitian operator
	into a trace-containing and traceless part,\cite{Kelly2012mapping}
	our strategy is not limited to electronic-state
	population differences of two-state systems but
	can also be applied in more general cases.

	The benefits associated with this simple strategy have been demonstrated on
	asymmetric spin-boson models with both Ohmic and Debye spectral densities. In each
	case, employing the aforementioned representation of the identity operator
	resulted in a significant improvement in the quality of results.
	Notably, in comparison to traditional approaches, which predict the wrong
	asymptotic limit for these systems,
	our approach consistently
	yields near quantitative accuracy in the cases investigated in this work.

	Note that in principle one can choose different approaches
	for each of the two correlation functions.
	It would therefore be possible to calculate a population difference
	using the ``double'' definition of $C_{\sigz\!\sigz}(t)$ and the ``single unity''
	$C_{\mathcal{I}\sigz}(t)$. For the systems presented
	here, the result is comparable in accuracy to ``double unity''.
	In future work, we will continue to test
	these strategies to discover the most reliable strategy.

	Different approaches have been previously proposed to address the
	shortcomings of classical mapping trajectories. For example,
	methods employing focused sampling \cite{Bonella2003mapping}
	force the total population to be unity. Note that these
	cannot therefore benefit further from our strategy.
	Another example is the symmetrical windowing approach,
	\cite{Cotton2013mapping,*Miller2016Faraday}
	in which binning is used to evaluate the effect of the state operators.
	Our approach is significantly simpler
	and has a stronger connection to the original LSC-IVR derivation
	as well as the formally exact properties of the mapping representation.

	The motivation underlying our strategy is similar to that behind
	work using generalized quantum master equations
	\cite{Kelly2015nonadiabatic,Kelly2016master,Reichman2016spinboson}.
	Both approaches use exact quantum mechanical information, where
	possible, to improve the accuracy, or rather avoid the loss in accuracy,
	of semiclassical trajectories.
	Notably, the accuracy gain which we demonstrate in this work is similar to what can be
	achieved using such generalized quantum master equation methods.

	Other quantum dynamics methods which utilize the mapping approach may also be
	able to benefit from employing a similar strategy to that suggested here.
	These include non-linearized semiclassical dynamics
	\cite{Stock1997mapping,Sun1997mapping,Thoss1999mapping,*Thoss2000pyrazine,Church2018MQCIVR},
	partially-linearized density matrix \cite{Huo2011densitymatrix},
	the forward-backward trajectory solution \cite{Hsieh2012FBTS,*Hsieh2013FBTS} of
	quantum-classical Liouville dynamics \cite{Kapral2015QCL}, and nonadiabatic
	ring-polymer molecular dynamics
	\cite{mapping,vibronic,Ananth2013MVRPMD,Duke2015MVRPMD,
	Hele2016Faraday,Chowdhury2017CSRPMD}.
	This study may thus have much wider implications
	for the methodology development of nonadiabatic dynamics simulations.

\section*{Acknowledgements}
\label{sec:acknow}
	\noindent
	M.A.C.S.~would like to acknowledge financial support through the ETH
	postdoctoral fellowship. The authors also acknowledge the support from the
	Swiss National Science Foundation through the NCCR MUST (Molecular Ultrafast
	Science and Technology) Network.

\clubpenalty10000
\bibliography{ref,references}

\end{document}